\documentclass{article}
\usepackage{spconf,amsmath,graphicx}
\usepackage{algorithm}
\usepackage{textcomp}
\usepackage{algpseudocode}
\usepackage{xcolor}
\usepackage{subcaption}
\usepackage{multirow}
\usepackage{pifont}
\usepackage{booktabs}
\usepackage{amsmath}
\usepackage{amsmath}  
\usepackage{amsfonts} 
\usepackage{amssymb}  
\usepackage{amsmath}
\usepackage{amsfonts}
\usepackage{algpseudocode}

\title{Optimum Power-Subcarrier Allocation and Time-Sharing in Multicarrier NOMA Uplink}

\name{Sagnik Bhattacharya, Kamyar Rajabalifardi, Muhammad Ahmed Mohsin, John M. Cioffi\thanks{Emails: \{sagnikb, kfardi, muahmed, cioffi\}@stanford.edu}}
\address{Dept. of Electrical Engineering, Stanford University, Stanford, CA, USA}
\usepackage{algpseudocode} 
\usepackage{algorithm} 

\begin{document}
%
\maketitle
\begin{abstract}
Currently used resource allocation methods for uplink multicarrier non-orthogonal multiple access (MC-NOMA) systems have multiple shortcomings. Current approaches either allocate the same power across all subcarriers to a user, or use heuristic-based near-far, strong channel-weak channel user grouping to assign the decoding order for successive interference cancellation (SIC). This paper proposes a novel optimal power-subcarrier allocation for uplink MC-NOMA. This new allocation achieves the optimal power-subcarrier allocation as well as the optimal SIC decoding order. Furthermore, the proposed method includes a time-sharing algorithm that dynamically alters the decoding orders of the participating users to achieve the required data rates, even in cases where any single decoding order fails to do so. Extensive experimental evaluations show that the new method achieves higher sum data rates and lower power consumption compared to current NOMA methods.
\end{abstract}

\section{Introduction}
\label{sec:intro}
Next-generation wireless communication applications include augmented / virtual reality (AR/VR), 3D gaming, multimodal artificial intelligence (MMAI), etc. These often require extremely high data rates ($\geq$ 500 Mbps per user) with low power consumption. This work investigates the case where a number of participating users communicate to the same base station (BS)/ access point (AP) on an uplink multiple access channel (MAC). This study can further be extended to downlink channels because of the established duality between the uplink multiple-access channel (MAC) and the downlink broadcast channel (BC). Traditional resource allocation methods use orthogonal multiple access (OMA) \cite{yin2006ofdma, seong2006optimal}, which allocates orthogonal frequency-time resource blocks to each user. OMA thus prevents separate user signals from interfering, and allows use of only a linear receiver to decode the user signal at the BS/AP. This OMA method predominates in Wi-Fi IEEE standards  (802.11b/g/n/ac/be) \cite{ieee80211ax2021} for resource allocation. However, as the number of participating users increases, the OMA methods fail to satisfy the required high data rates and provide suboptimal rates. This is because the OMA methods do not use to their advantage the interference caused by the signals of other users, if more than one user are co-allocated same frequency-time resource blocks. 
To address the limitations of OMA, non-orthogonal multiple access (NOMA) resource allocation techniques \cite{noma6, tse_viswanath_2005, islam2018resource, zhang2018energy} have been introduced. NOMA enables resource sharing through successive interference cancellation (SIC) at the BS/AP, where the decoding order is crucial. As each user is decoded, interference from users earlier in the decoding order is subtracted, while remaining interference is treated as noise alongside the additive white Gaussian noise (AWGN).
Optimized NOMA outperforms OMA systems. \cite{6692652} explores NOMA in SISO scenarios, where SIC's optimal decoding order is based on channel strength. However, in contrast to the SISO case, for multi-user MIMO, no inherent decoding order exists based on users' channels \cite{7938651}, since the channel corresponding to each user is a vector. 
Several heuristic approaches for MIMO decoding order, such as limiting to two users \cite{noma8} or clustering users \cite{noma9} as near user-far user, often lead to suboptimal data rates and power consumption. These methods assume that all users share the entire bandwidth, relying solely on power or code domain multiple access, which limits performance. In contrast, \cite{noma11} optimizes power and subcarrier allocation in MC-NOMA, but assumes a MISO system with a single antenna at the AP, and a heuristic decoding based on channel coefficients. This heuristic decoding order derivation is suboptimal for MU-MIMO because it does not generalize to vector channels. Moreover, \cite{noma11} employs a distinct decoding order for each subcarrier, which leads to increased latency during the detection process. This is primarily because most current communication systems use low density parity-check codes (LDPC) \cite{gallager1962low}, and having a different decoding order for each subcarrier increases the decoder window of calculation. Further, a different decoding order for each subcarrier is in itself, suboptimal.

In this paper, we propose a novel algorithm for optimal power-subcarrier allocation and decoding order for uplink MC-NOMA. It eliminates latency-inducing per-subcarrier decoding order by optimizing a common order across all subcarriers, avoiding heuristic assumptions. By decoupling power-subcarrier allocation and decoding order derivations, the proposed method preserves convexity, ensuring efficient global optimum achievability. This contrasts with non-convex power allocation problems in prior work \cite{noma5, noma6, noma7, noma8}. Additionally, a novel time-sharing algorithm adaptively switches between multiple decoding orders to achieve higher data rates than a single order. This method surpasses the gains from using different decoding orders on different subcarriers and achieves data rates that fixed decoding orders, as in \cite{noma11}, cannot. The combination of optimal power allocation, derived decoding order, and adaptive time-sharing meets data rate and power consumption requirements that current OMA/MC-NOMA methods cannot. Extensive experiments demonstrate that the proposed algorithm outperforms OMA, NOMA, and MC-NOMA by 39\%, 28\%, and 16\% in data rates under various SNR conditions.


\section{System Model}
\label{sec:system-model}


We consider an uplink multiple access channel (MAC), where users transmit symbols to an AP. Specifically, there are $U$ users, with the $u^{\textrm{th}}$ user having $L_{x,u}$ antennas and transmitting a symbol vector $\mathbf{x}_u \in \mathbb{C}^{L_{x,u}}$ to the AP. The AP has $L_y$ antennas. Hence, the AP's received signal $\mathbf{y}$ is:

\begin{equation}
\mathbf{y} = \mathbf{H} \cdot \mathbf{x} + \mathbf{n}
\end{equation}

where $\mathbf{x} \in \mathbb{C}^{L_x}$ represents the concatenation of all symbols $x_u$ sent by the users, and $\mathbf{y} \in \mathbb{C}^{L_y}$. Here, $L_x = \sum_{u=1}^{U} L_{x,u}$. Furthermore, $\mathbf{H} \in \mathbb{C}^{L_y \times L_x}$ represents the channel matrix, and $\mathbf{n} \in \mathbb{C}^{L_y}$ is the additive white gaussian noise (AWGN) at the receiver.

In a multi-carrier system with $N$ subcarriers, Equation (1) applies independently to each subcarrier. The receiver employs successive interference cancellation (SIC). This method is optimal because, given a specific user power allocation, an optimal SIC decoding order can achieve the information-theoretic sum capacity bound, provided good codes are also in use.

Let $\pi$ represent the decoding order vector, where $\pi(1)$ corresponds to the decoding order for user 1. Similarly, $\pi^{-1}$ is defined as the inverse of $\pi$, with $\pi^{-1}(1)$ indicating the user that is decoded first. Using this, the achievable data rate for  user $\pi^{-1}(u)$ is:
\begin{equation}
\begin{aligned}
&b_{\pi^{-1}}(u) = \sum_{i=u}^{U} b_{\pi^{-1}}(i) - \sum_{i=u+1}^{U} b_{\pi^{-1}}(i) \\
&\small{= \log_2 \left( \left| \frac{\mathbf{R}_{\textrm{noise}} + \sum_{i=u}^{U} \mathbf{H}_{\pi^{-1}(i)} \cdot \mathbf{R}_{\boldsymbol{x}\boldsymbol{x}}(\pi^{-1}(i)) \cdot \mathbf{H}^*_{\pi^{-1}(i)}}{\mathbf{R}_{\textrm{noise}} + \sum_{i=u+1}^{U} \mathbf{H}_{\pi^{-1}(i)} \cdot \mathbf{R}_{\boldsymbol{x}\boldsymbol{x}}(\pi^{-1}(i)) \cdot \mathbf{H}^*_{\pi^{-1}(i)}} \right| \right)}
\end{aligned}
\end{equation}

Here, $\mathbf{R}_{\textrm{noise}}$ represents the noise auto-correlation matrix, while $\mathbf{H}_{\pi^{-1}(u)}$ stands for the channel matrix between the AP and user $\pi^{-1}(u)$. Equation~(2) is derived under SIC assumption, where each user cancels interference from the symbols of users that have been decoded earlier in the sequence and treats the signals of users decoded later as noise. This corresponds to the well-known chain rule of information theory. Finding the optimal decoding order $\pi$ among $U!$ possible combinations is a challenging task, especially for vector channels, without making heuristic assumptions, which this work addresses in Section~\ref{sec:optimal-decoding-order}.

\section{Optimal Power-Subcarrier Allocation}

The optimal power and subcarrier allocation is framed as a primal-dual optimization problem. The primal problem minimizes the total user energy (or equivalently, the power), subject to the minimum data rate requirements. On the other hand, the dual problem maximizes the total user's data rate sum while adhering to the energy constraints. The weighted-sum energy minimization problem is:

\begin{equation}
\begin{aligned}
&\min_{\left\{\mathbf{R}_{\boldsymbol{x} \boldsymbol{x}}{(u, n)}\right\}} \sum_{u=1}^{U} \sum_{n=1}^{N} w_u \cdot \operatorname{trace}\left(\mathbf{R}_{\boldsymbol{x} \boldsymbol{x}}(u, n)\right) \\
&\text{C1}: \mathbf{b}=\sum_{n=1}^{N}\left[b_{1, n}, b_{2, n}, \ldots ,b_{U, n}\right]^T \succeq \mathbf{b}_{\min } \succeq \mathbf{0} \\
&\text{C2}: \footnotesize{\sum_{u \in T}\sum_{n=1}^{N} b_{u, n} \leq \log_2\left| \frac{\mathbf{R}_{\textrm{noise}} + \sum_{u \in T} \mathbf{H}_{u, n} \cdot \mathbf{R}_{\boldsymbol{x}\boldsymbol{x}}(u, n) \cdot \mathbf{H}_{u, n}^*}{\mathbf{R}_{\textrm{noise}}} \right|} \\
&\small{\quad \forall T \subseteq \{1, 2, \cdots, U\}} \\
&\text{C3}: \mathbf{R}_{\boldsymbol{x}\boldsymbol{x}}(u,n) \succcurlyeq \mathbf{0} \quad \forall u \in \{1, 2, ..., U\},  \forall n \in \{1, 2, ..., N\}
\end{aligned}
\end{equation}

In this formulation, \( \mathbf{R}_{\boldsymbol{xx}}(u,n) \in \mathbb{R}^{L_{x,u}\times L_{x,u}} \) represents the autocorrelation matrix of \( \boldsymbol{x} \) for user \( u \) on the \( n^{th} \) subcarrier, and \( w_u \) is the non-negative weight that prioritizes the energy contribution of the \( u^{th} \) user in the minimization. The term \( b_{u,n} \) is the data rate achieved by the \( u^{th} \) user on the \( n^{th} \) subcarrier, while \( \mathbf{b}_{\min} \) is a vector of length \( U \), indicating the minimum required data rates for each user. Constraint C1 enforces the minimum data rate requirement, and C2 ensures that for any subset of users, the sum of their data rates is less than or equal to the theoretical capacity limit. Finally, C3 ensures that the autocorrelation matrix \( \mathbf{R}_{\boldsymbol{x}\boldsymbol{x}}(u,n) \) is positive semi-definite, meaning it must be symmetric and all its eigenvalues must be non-negative. The weights \( w_u \) determine the relative importance of minimizing each user's energy consumption.

This optimization problem yields the optimal power allocations for the users across subcarriers to meet the desired data rates. Notably, this optimization is independent of the SIC decoding order and is a convex optimization problem. By separating the derivation of the optimal power and subcarrier allocations from the decoding order, convexity is preserved, allowing for an efficient and optimal solution to the problem. This is in contrast to the non-convex power allocation problems in related work \cite{noma5, noma6, noma7, noma8}, which arise from heuristic assumptions about the decoding order. With the optimal power allocations determined for each subcarrier, we proceed to derive the optimal decoding order.

\section{Optimal Decoding Order Derivation}
\label{sec:optimal-decoding-order}

Since the heuristic decoding order based on channel state information (CSI) results in suboptimal power allocation solutions, the proposed method derives the optimal decoding order over all subcarriers.
The optimal decoding order's determination begins by addressing the dual problem of Equation~\eqref{eq:energy-sum-minimize}, which involves maximizing the data-rate sum  under a maximum energy budget constraint. The dual problem is:

\begin{equation}
\begin{aligned}
\label{eq:energy-sum-minimize}
&\max_{\left\{R_{\boldsymbol{x} \boldsymbol{x}}(u, n)\right\}} \quad \sum_{u=1}^{U} \theta_u \cdot \left(\sum_{n=0}^{N} b_{u, n}\right) \\
&\text{s.t.} \\
&\text{C1}: \mathcal{E} = \sum_{n=0}^{N} \left[\mathcal{E}_{1, n}, \mathcal{E}_{2, n}, \ldots, \mathcal{E}_{U, n}\right]^T \preceq \boldsymbol{\mathcal{E}}_{\max} \\
&\text{C2}: \sum_{u \in T}\sum_{n=1}^{N} b_{u, n} \leq \log_2\left| \frac{\mathbf{R}_{\textrm{noise}} + \sum_{u \in T} \mathbf{H}_{u, n} \mathbf{R}_{\boldsymbol{x}\boldsymbol{x}}(u, n) \mathbf{H}_{u, n}^*}{\mathbf{R}_{\textrm{noise}}} \right|, \\
&\quad \forall T \subseteq \{1, \ldots, U\} \\
&\text{C3}: \mathbf{R}_{\boldsymbol{x}\boldsymbol{x}}(u,n) \succcurlyeq \mathbf{0}, \; \forall u \in \{1, \ldots, U\}, \; \forall n \in \{1, \ldots, N\}
\end{aligned}
\end{equation}

Here, $\mathcal{E}_{u, n} = \textrm{trace}\left(\mathbf{R}_{\boldsymbol{xx}}(u,n)\right)$ represents the energy assigned to the $u^{th}$ user on the $n^{th}$ subcarrier, and $\theta_{u}$ denotes the non-negative weights applied to the user data rates. Notably, $\theta_{u}$ (for all $u \in {1, 2, \cdots, U}$) are dual variables associated with the minimum required data rate constraints in the primal problem given by Equation~\eqref{eq:energy-sum-minimize}. As established in \cite{book}, the maximum weighted rate sum is always achievable. Furthermore, it is proven that, for this optimal data rate sum, the decoding order $\mathbf{\pi}$ must follow the inequality based on the dual variables $\theta_{u}$:

\begin{equation}
\label{eq:decoding_order}
\theta_{\mathbf{\pi}^{-1}(U)} \geq \theta_{\mathbf{\pi}^{-1}(U-1)} \geq … \geq \theta_{\mathbf{\pi}^{-1}(1)}
\end{equation}

Consequently, the optimal decoding order is determined by the ranking of the dual variables corresponding to the minimum required data rates in the energy sum minimization problem. Solution of this convex optimization problem using a primal-dual approach, as outlined in Equation~\eqref{eq:energy-sum-minimize}, automatically obtains the dual variables $\theta$, and thus the optimal decoding order.



\section{Time Sharing}
\label{sec:time-sharing}



A strict ordering of the $\theta_u$ for all $u \in \{1, 2, ..., U\}$ ensures a unique decoding order, necessary for achieving required data rates. However, Equation~\eqref{eq:decoding_order} does not guarantee strict inequalities, leading to cases where equal $\theta_u$ values mean no unique decoding order can meet the data rates \cite{book}. Existing NOMA approaches with heuristic single decoding orders may fail under these conditions. The proposed method addresses this issue, through a novel time-sharing approach that combines multiple decoding orders to achieve the required data rates when any single decoding order is insufficient. This time-sharing algorithm handles cases where $\theta_u$ values are identical by grouping users with the same $\theta_u$ into clusters. Within each cluster, all possible permutations of the users are considered as potential decoding orders.

\begin{table}[t!]
\centering
\caption{ Time Sharing Transmit Power Consumption Reduction, Noise Power -65 dBm}
\resizebox{0.49\textwidth}{!}{ 
\begin{tabular}{|c|c|c|c|c|c|c|c|c|c|c|}
\hline
\textbf{Time Sharing} & \multicolumn{3}{|c|}{\textbf{Transmit}} & \multicolumn{3}{|c|}{\textbf{Data Rates}} & \textbf{Time-shared} & \multicolumn{3}{|c|}{\textbf{SIC Decoding}} \\ 
 &  \multicolumn{3}{|c|}{\textbf{Power (dBm)}} &  \multicolumn{3}{|c|}{\textbf{(Mbps)}} & \textbf{Fraction} & \multicolumn{3}{|c|}{\textbf{Order}} \\ 
\hline
\multirow{3}{*}{\checkmark} & \multirow{3}{*}{15} & \multirow{3}{*}{14.3} & \multirow{3}{*}{15} & 398.01 & 470.48 & 632.23 & 0.52 & 3 & 2 & 1 \\
& &  &  & 691.78 & 242.32 & 565.91 & 0.17 & 1 & 3 & 2 \\
& &  &  & 565.91 & 691.78 & 242.32 & 0.31 & 2 & 1 & 3 \\ \hline 
\ding{55} & 15.8 & 16 & 15.4 & 500 & 500 & 500 & 1.00 & 3 & 2 & 1 \\
\hline
\end{tabular}
}\label{tab:time-sharing2}
\end{table}

Each cluster acts as a composite user, and these along with single users are ordered by strict inequalities. Considering all permutations within each cluster generates several candidate decoding orders, $num\_ord$, all requiring the same power allocation \cite{book}. The time-sharing algorithm evaluates each decoding order using SIC with optimal power allocation, denoted as $s_i$. It then computes a linear combination of these rates using linear programming to meet the required data rates. Time-sharing weights ($t\_w$) represent the usage fractions for each order, ensuring the required rates are met when a single order is insufficient. The linear programming formulation is:


\begin{align*}
\text{minimize} \quad & \sum_{i=1}^{\text{$num\_ord$}} z_i \\
\text{subject to} \quad & \text{C1: } z_i \in \{0, 1\} \quad \forall i \in \{1, \dots, \text{$num\_ord$}\} \\
& \text{C2: } t\_{w_i} \le z_i \quad \forall i \in \{1, \dots, \text{$num\_ord$}\} \\
& \text{C3: } t\_{w_i} \ge 0 \quad \forall i \in \{1, \dots, \text{$num\_ord$}\} \\
& \text{C4: } \sum_{i=1}^{\text{$num\_ord$}} t\_{w_i} = 1 \\
& \text{C5: } \sum_{i=1}^{\text{$num\_ord$}} s_i \cdot t\_{w_i} = \mathbf{b}_{\min }
\end{align*}

This formulation introduces a binary array $z$ of length $num\_ord$. The objective function and constraint C2 work together to minimize the number of non-zero time-sharing coefficients, effectively reducing the number of decoding orders used in time-sharing. This is crucial because switching decoding orders during time-sharing introduces latency, and minimizing the number of switches reduces this latency. Constraints C3 and C4 ensure that the time-sharing coefficients are non-negative and sum to 1, as they represent the fraction of time each decoding order is applied. Finally, constraint C5 ensures that the time-shared average data rates are equal to the required user data rates.

To demonstrate the time-sharing scenario, Table~\ref{tab:time-sharing2} presents an example. The objective is to achieve target data rates of 500 Mbps per user for a low-rank channel with two AP antennas and three users, each with one antenna. The users are positioned equidistantly from the AP at 3 meters, and the results of the proposed algorithm for both optimal power-subcarrier allocation and time-sharing. The required transmit power for the three users without time-sharing is \{15.8, 16, 15.4\} dBm, whereas with time-sharing, the power levels are reduced to \{15, 14.3, 15\} dBm. As shown in the table, time-sharing offers the advantage of dynamically varying the decoding order in SIC for different time fractions, achieving the required data rates on average with reduced power consumption.

\section{Performance Evaluation}
\label{sec:performance}
\begin{figure}[t]
    \centerline{
        \includegraphics[trim = {0, 200, 0, 210}, clip, height = 5.7cm]{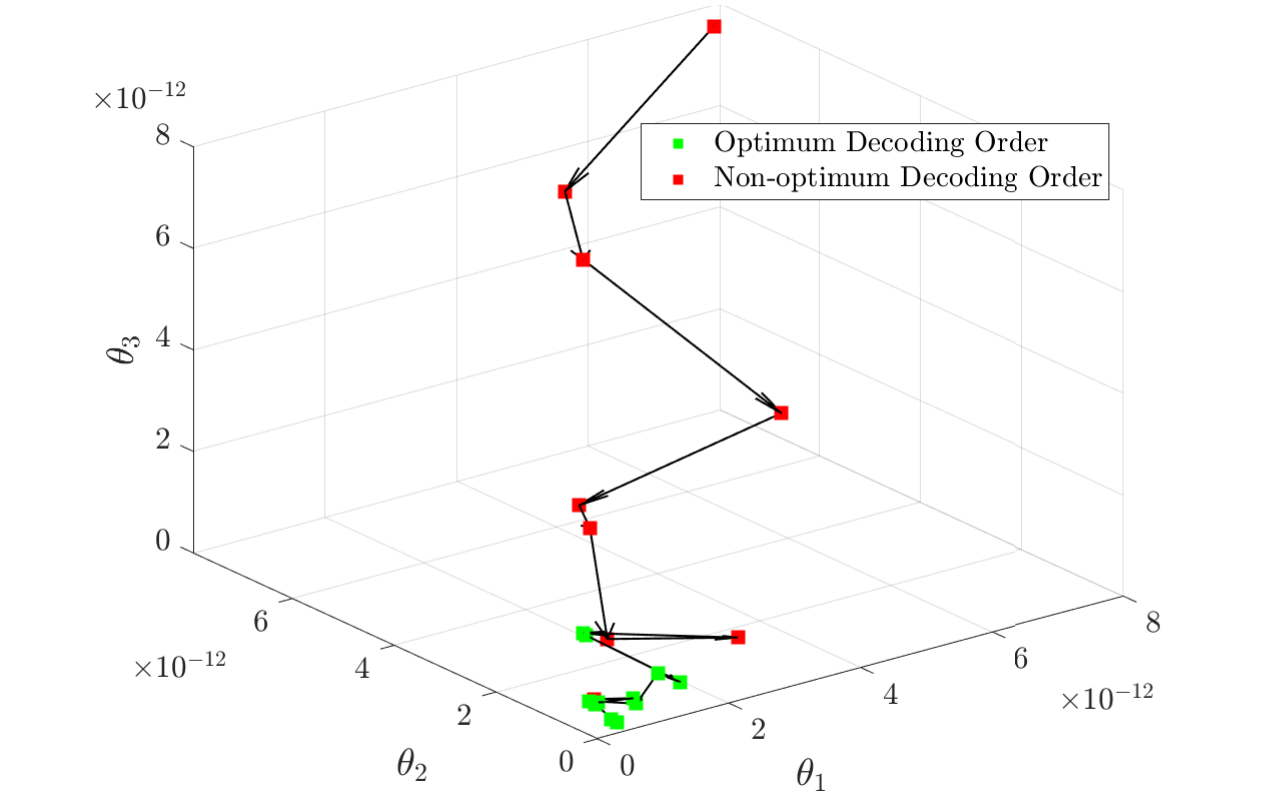}}
    \caption{Values of Lagrange multipliers per iteration of ellipsoid algorithm, 3 users with 3m distance from AP}
    \label{fig:decoding_order}
\end{figure}

\begin{figure}[t]
    \centerline{
        \includegraphics[width=0.45\textwidth]{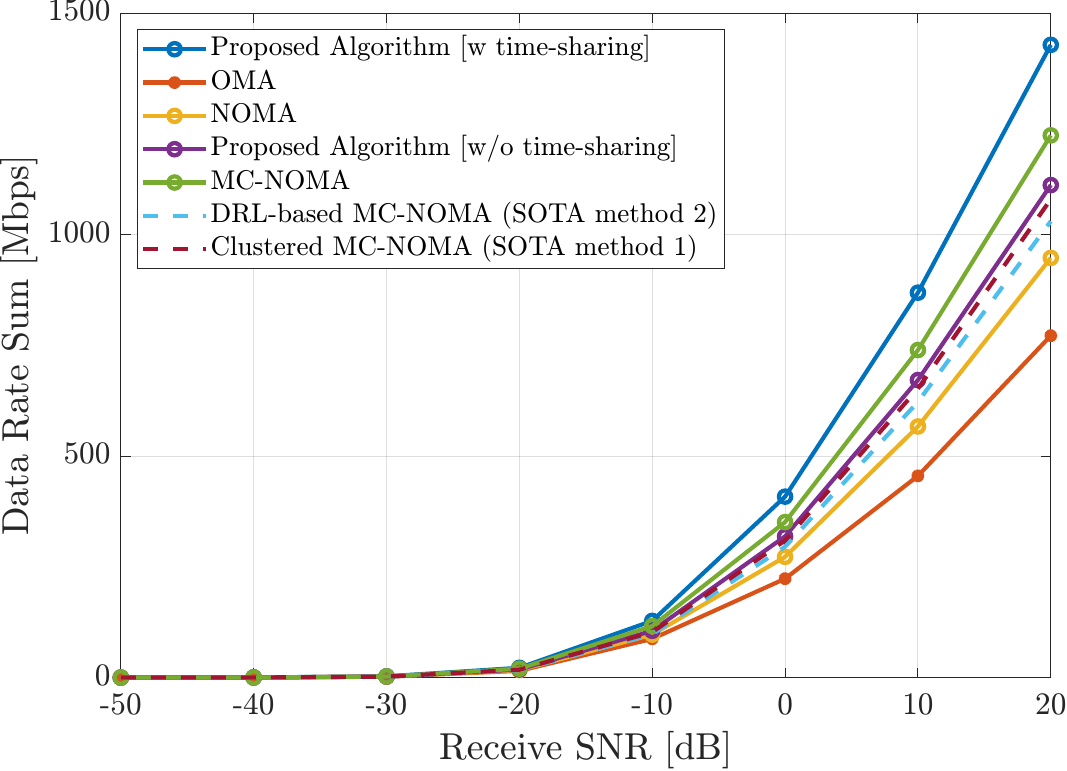}}
    \caption{Sum Rate versus receive SNR for Single Antenna Per User with distances from AP=\{3m, 3m, 3m\}: Baselines and Proposed Algorithm}
    \label{fig:sumrate}
\end{figure}

\textbf{Heuristic decoding for vector channels:}
Fig.~\ref{fig:decoding_order} represents convergence of the dual variables $\theta_i$ as the convex optimization algorithm in Eq.~\eqref{eq:energy-sum-minimize} converges. The decoding order, $\pi$ is determined by sorted order of $\theta_i$, following the principle that users with higher $\theta$ values are decoded later, as shown in Eq.~\eqref{eq:decoding_order}. This approach, derived from the dual variables of the energy minimization problem (Equation~\eqref{eq:energy-sum-minimize}), provides an efficient method for determining optimal decoding orders in multi-user vector channels, unlike current work, which assumes heuristic decoding order.\\ 
\textbf{Sumrate vs. SNR:} In Fig.~\ref{fig:sumrate}, we observe the sumrate against SNRs under different scenarios. The proposed algorithm with time-sharing demonstrates highest sumrate and the difference increases, especially for high SNRs. The proposed algorithm without time-sharing still performs better than baseline methods NOMA and OMA. Our proposed power-subcarrier allocation algorithm, coupled with the proposed time-sharing mechanism performs better than baseline methods~\cite{7938651, yin2006ofdma, tse_viswanath_2005} for all SNRs.

\section{Conclusion}
In this paper, we propose a novel optimum power subcarrier allocation and optimum decoding order derivation for MC-NOMA uplink. We also introduce a novel time sharing algorithm, which helps achieve the required data sum rates with the same decoding order across all subcarriers, preserving latency. We show via simulations that the novel power-subcarrier allocation algorithm, coupled with the time sharing mechanism outperforms current resource allocation methods for MC-NOMA. 

\bibliographystyle{IEEEbib}
\bibliography{strings,refs}

\end{document}